\documentclass[twocolumn]{aastex63}
\usepackage{orcidlink}

\usepackage{hyperref}
\hypersetup{
  colorlinks   = true, 
  linkcolor    = red, 
  citecolor   = blue 
}

\newcommand{\beq}{\begin{equation}}
\newcommand{\eeq}{\end{equation}}
\newcommand{\beqn}{\begin{eqnarray}}
\newcommand{\eeqn}{\end{eqnarray}}

\newcommand{\lppr}{\stackrel{<}{\scriptstyle \sim}}

\accepted{July 2025}
\submitjournal{ApJ}

\shorttitle{Synchrotron-limited Electron Acceleration in Shearing Flows}
\shortauthors{Rieger and Duffy}

\begin{document}
\title{Synchrotron-limited Particle Acceleration in Relativistic Shearing Flows}

\correspondingauthor{Frank M. Rieger}
\email{frank.rieger@ipp.mpg.de}

\author{Frank M. Rieger \orcidlink{0000-0003-1334-2993}}
\affiliation{Max Planck Institute for Plasma Physics (IPP), Boltzmannstraße 2, 85748 Garching, Germany}
\affiliation{Institute for Theoretical Physics, University of Heidelberg, \\
Philosophenweg 12, 69120 Heidelberg, Germany}

\author{Peter Duffy}
\affiliation{School of Physics, University College Dublin, Belfield, Dublin 4, Ireland}

\begin{abstract}
Fermi-type shear particle acceleration is a promising mechanism for sustaining 
ultra-relativistic particles along the kilo-parsec scale jets in Active Galactic 
Nuclei (AGNs). We explore the possibility of synchrotron-limited electron 
acceleration in mildly relativistic shearing flows and present numerical solutions 
to the corresponding particle transport equation. We compare our findings with 
analytical calculations to infer an effective electron cutoff energy, and discuss 
the relationship to a simplified box model treatment.
The results show that mildly relativistic large-scale jets offer a suitable 
environment for distributed electron acceleration beyond Lorentz factors of 
$\gamma_e \sim 10^8$.
\end{abstract}

\section{Introduction}
Fast shearing flows are known to be conducive to efficient Fermi-type particle 
acceleration \citep{Rieger2019,Lemoine2019}. The relativistic jets and
winds in AGNs are sites where shear particle acceleration is 
thought to be efficient in producing and maintaining ultra-relativistic particles 
\citep[][]{Webb2018,Webb2023,Rieger2019b,Rieger2021,
Tavecchio2021,Merten2021,Seo2023,Wang2021,Wang2024}. 

The field has seen some significant developments over the years \citep[]
[for review]{Rieger2019}. Early modeling on particle transport in shearing flows 
found that in the absence of radiative losses, the accelerated particle 
(momentum) distribution is a power-law spectrum, $f(p) 
\propto p^{-s}$, with spectral index $s=(3+\alpha)$ for a mean scattering time 
scaling as $\tau \propto p^{\alpha}$ \citep{Berezhko1981,Berezhko1982,Rieger2006}. 
However, this index only applies to highly relativistic
($\beta\rightarrow1$) flows, while for sub-relativistic speeds  much softer spectra 
are encountered \citep{Webb2018,Webb2019,Rieger2019b} as the momentum gain 
for a particle scattered across the flow competes with diffusive escape from the system 
\citep{Jokipii1990,Rieger2006}. At mildly 
relativistic flow speeds, the particle momentum distribution $f(p)$ becomes 
sensitive to the underlying flow profile \citep{Rieger2022}, providing 
an additional observational diagnostic.
For electrons radiative losses usually become important at 
high energies. In particular, in the box model synchrotron losses are expected 
to lead to an exponentially shaped cut-off in the electron distribution at the 
energy scale where acceleration is balanced by cooling \citep{Liu2017,Wang2021}. 
An in-depth exploration of this appears relevant as shear particle acceleration 
represents one of the most promising, distributed mechanisms  for sustaining 
ultra-relativistic electrons (with Lorentz factors up to $\gamma_e \sim 10^8$) 
along the kilo-parsec scale jets in AGNs as implied by the detection of 
extended X-ray synchrotron and inverse-Compton (IC) VHE emission 
\citep[e.g.,][]{Sun2018,Perlman2020,HESS2020,Meyer2023IAUS,Breiding2023,He2023}. 
In the following, we focus on solutions of the generalized particle transport 
equation in the presence of synchrotron losses, and discuss their relevance 
in the context of a (leaky) box model treatment.

\section{A simplified box treatment}\label{leaky_box}
It has been shown that a box model for the spatially averaged, steady-state momentum 
phase space distribution $f(p)$ captures the essential features of shear acceleration 
\cite[][]{Liu2017,Rieger2019b,Wang2021}.
With a momentum diffusion coefficient $D_p$, escape timescale $\tau_{\rm esc}$,
synchrotron loss coefficient $\tilde{\beta}_s \equiv 4 B^2 e^4/(9m^4c^6)$ and 
source term $Q(p)$, the basic equation is 
\beq\label{diffusion}
\frac{1}{p^2} \frac{\partial}{\partial p }
\left(p^2 D_p \frac{\partial f}{\partial p}\right)
+\frac{1}{p^2} \frac{\partial}{\partial p }
\left(\tilde{\beta}_s p^4 f\right) 
- \frac{f}{\tau_{\rm esc}}+ Q(p) = 0 \,.
\eeq 
For a momentum-dependent particle mean free path, $\lambda = c\tau 
\propto p^\alpha$, the momentum diffusion coefficient $D_p = D_0 \,p^{2+\alpha}$ 
gives a characteristic acceleration timescale that decreases with momentum, 
$t_{\rm acc}(p)=p^2/([4+\alpha] D_p)\propto p^{-\alpha}$, as higher momentum 
particles are scattered across larger parts of the shear flow. 
The same scattering gives also rise to a spatial diffusion coefficient 
$\kappa(p)\propto \lambda\propto p^\alpha$ and a corresponding escape 
timescale $\tau_{\rm esc}=(\Delta r)^2/2 \kappa(p)\propto p^{-\alpha}$ where 
$\Delta r$ is the width of the shear region. That the acceleration and escape 
timescales have the same momentum dependence is a basic requirement of 
any Fermi-type acceleration process to result in a power law particle distribution. 
If synchrotron losses are negligible (i.e., for $\tilde{\beta}_s \rightarrow 0$), 
such as usually the case for cosmic-ray protons, using $f\propto p^{-s}$ in 
equation (\ref{diffusion}) yields a power-law index 
\beq\label{s_LB}
 s = \frac{(3+\alpha)}{2} + \sqrt{\frac{(3+\alpha)^2}{4} 
    + (4+\alpha)\, \frac{t_{\rm acc}}{\tau_{\rm esc}}}\,,
\eeq
recovering the limit of $s=3+\alpha$ when $t_{\rm acc}\ll \tau_{\rm esc}$ 
\citep{Rieger2019b}. On the other hand, taking synchrotron losses for
electrons into account typically introduces (for $\alpha<1$) an exponentially 
shaped cut-off
\beq\label{shape}
f(p) \propto \exp\left[-\frac{4+\alpha}{1-\alpha} 
                   \left(\frac{p}{p_{\max}}\right)^{1-\alpha}\right]
\eeq at the momentum scale $p_{\max} =\gamma_{\rm max} m_e c = 
[(4+\alpha)D_0/\tilde{\beta}_s]^{1/(1-\alpha)}$ where acceleration balances 
cooling, i.e., when $t_{\rm acc}(p) = t_{\rm cool}(p)\equiv 
\frac{1}{\tilde{\beta}_s p}$ \citep{Liu2017,Wang2021}.

\section{Generalized Particle Transport}
The (spatially averaged) box formulation of Sec.~\ref{leaky_box} can be compared 
for accuracy with a generalized (space-dependent) particle transport description. 
This comparison is essential to understand how the conventional box model, commonly 
employed in phenomenological studies, must be adjusted to suitably accommodate 
spatial variations. Our analysis below reveals that this has a particular impact on the 
inferred acceleration timescales, and thus on maximum achievable particle energies.
As shown by \citet{Webb1989}, the evolution of the phase space distribution of 
energetic charged particles in relativistic flows can be cast in a diffusive 
particle transport equation. 
For a relativistic jet along the z-direction with cylindrical shear flow profile 
$\beta(r)=u_z(r)/c$, the corresponding, steady-state particle transport equation 
for the $z$-integrated, phase-space distribution function $f(r,p)$ in the presence 
of electron synchrotron losses is \citep[][]{Webb2018}
\beqn\label{pte}
\frac{1}{p^2}
\frac{\partial}{\partial p} \left(p^4 \Gamma_s \tau 
\frac{\partial f}{\partial p}\right) 
&+&\frac{1}{r} \frac{\partial}{\partial r} 
\left(\kappa r \frac{\partial f}{\partial r}\right) 
\nonumber \\
&+& \frac{1}{p^2} \frac{\partial}{\partial p} \left(\tilde{\beta}_s p^4 f\right)
+ Q(r,p)=0\,,\quad
\eeqn where $Q(r,p)$ represents the source term, $\tau = \lambda/c$ is the mean 
scattering time, $\kappa =c^2 \tau / 3$ defines the spatial diffusion coefficient, 
 $\tilde{\beta}_s$ is the synchrotron loss coefficient, and 
\beq\label{GammaS}
\Gamma_s \equiv \Gamma_s(r) = \frac{c^2}{15}\, 
\gamma_b(r)^4 \left(\frac{d\beta}{dr}\right)^2
\eeq denotes the relativistic shear flow coefficient, with flow Lorentz factor 
given by $\gamma_b(r)^2 = 1/(1-\beta(r)^2)$\citep{Rieger2006,Webb2018}. 
Formally, equation~(\ref{pte}) represents a mixed-frame particle transport 
equation, in which the momentum variable $p$ is evaluated in the co-moving 
flow frame\footnote{In what follows, this variable (and similarly the scattering
time and synchrotron loss coefficient) is understood to refer to the comoving frame. 
For simplicity, we drop  the 'prime' notation.} and the spatial coordinate $r$ 
measured in the laboratory frame.
The latter is achieved by a Lorentz boost from the local fluid to the 
laboratory frame. This approach is motivated by the fact that the scattering 
process, i.e. the collision operator, is most conveniently evaluated in the 
local fluid frame \citep[][]{Kirk1988}. In general, the local scattering 
time $\tau$ can be a function of both, space $r$ and momentum $p$. For 
convenience and to facilitate comparison, we consider in the following 
$\tau$ to be solely dependent on $p$, i.e., $\tau \equiv \tau(p) = \tau_0\, 
p^{\alpha}$, $\alpha>0$. In the quasi-linear limit $\alpha$ is related to the 
power index $q$ of the turbulence spectrum by $\alpha=2-q$ 
\citep[][]{Liu2017}. In practice, we employ a Kolmogorov ($q=5/3$) scaling 
$\lambda = c \tau = \xi^{-1} \Lambda_{\rm max} (r_g/\Lambda_{\rm max})^{1/3}$ 
where $r_g = \gamma m_e c^2/(eB)$ is the electron gyro-radius, $\xi \leq 1$ 
and $\Lambda_{\rm max}$ is the turbulence coherence length \citep[][]{Rieger2019b}.
We note that a Kolmogorov-type scaling is supported by MHD jet simulations 
\citep{Wang2023}. Unless otherwise stated, we choose $\xi=0.2$.

Since, in the absence of radiative losses, the shape of the accelerated particle 
distribution is sensitive to the underlying flow profile \citep[][]{Rieger2022}, 
we study the impact of synchrotron losses by numerically solving the particle 
transport equation, equation~(\ref{pte}), for two different cases. Firstly, a 
linearly decreasing velocity profile 
\beq\label{linear}
 \beta_l(r) = \beta_0 \left(1- (r/r_2)\right)\,,
\eeq with on-axis speed $\beta_0$ and outer radius $r_2$, 
and secondly a power-law velocity profile, with $b\ge 1$, 
\beq\label{power-law}
\beta_p(r) = \frac{\beta_0}{(1+a_p^2\, [r/r_2]^2)^{b/2}}\,.
\eeq 
Formally, the linearly decreasing profile, equation~(\ref{linear}), results 
in a monotonically decreasing $\Gamma_s$, while the power-law one yields a 
$\Gamma_s$, that first rises and then subsequently decreases monotonically 
\citep[][]{Rieger2022}.\\
In the following we present numerical solutions of equation~(\ref{pte}) for 
a mono-energetic source term  with $p=p_0 =\gamma_0 m_e c$ at $r=0.1\,r_2$, 
using a finite element method, and with a set of parameters appropriate 
for mildly relativistic ($\gamma_b \lppr 4$) large-scale AGN jets; where 
$B=10~\mu$G, $\Lambda_{\rm max} \simeq r_2 = 100$ pc, $\gamma_0=10^6$ 
unless otherwise stated \citep[][]{HESS2020,Wang2021}. An inner boundary 
condition $\partial f/\partial r \rightarrow 0$ as $r \rightarrow 0$ has 
been applied throughout.

\subsection{Numerical solutions for linear profiles}\label{subsec31}
\begin{figure}[htbp]
\begin{center}
\includegraphics[width = 0.48 \textwidth]{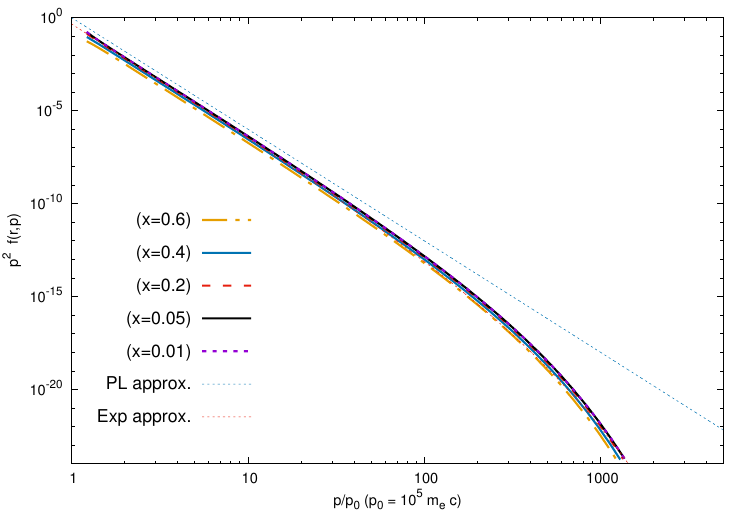}
\caption{The accelerated electron momentum distribution $n(r,p) \propto p^2 f(r,p)$ 
as a function of $p/p_0$ for a linearly decreasing flow profile, equation~(\ref{linear}), 
with $\beta_0 = 0.7$, evaluated at different radii $r=x\,r_2$. A spatially constant 
$\tau \propto p^{\alpha}$ with $\alpha=1/3$, and mono-energetic injection with 
$\gamma_0=10^5$ have been assumed. Since the flow Lorentz factor $\gamma_b$ is 
everywhere close to one, the synchrotron cut-offs are close to each other. The 
thin blue-dotted line shows the corresponding power-law (PL) solution ($s\simeq 8$) 
in the absence of synchrotron losses \citep{Rieger2022}, while the thin red-dotted 
curve (Exp) gives the one multiplied by $\exp(-0.09\,[p/p_0]^{2/3})$.}
\label{fig1}
\end{center}
\end{figure}

\begin{figure}[htbp]
\begin{center}
\includegraphics[width = 0.48 \textwidth]{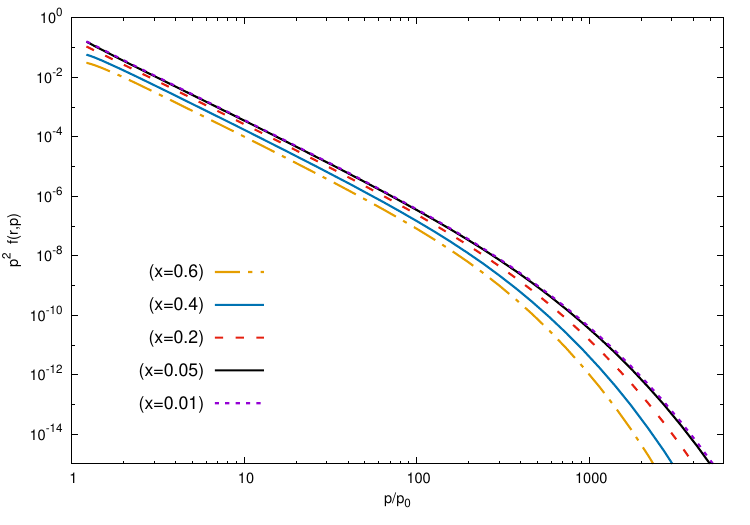}
\caption{The accelerated electron momentum distribution $n(r,p) \propto p^2 f(r,p)$ 
for a linearly decreasing flow profile, equation~(\ref{linear}), with $\beta_0 = 0.95$, 
evaluated at different radii $r=x\,r_2$. A spatially constant $\tau \propto p^{\alpha}$ 
with $\alpha=1/3$ and $\gamma_0 = 10^6$ have been assumed. The distribution 
extends to higher momenta at smaller radii where faster flow speeds ($\gamma_b 
> 1$) are met.}
\label{fig2}
\end{center}
\end{figure}

\begin{figure}[htbp]
\begin{center}
\includegraphics[width = 0.48 \textwidth]{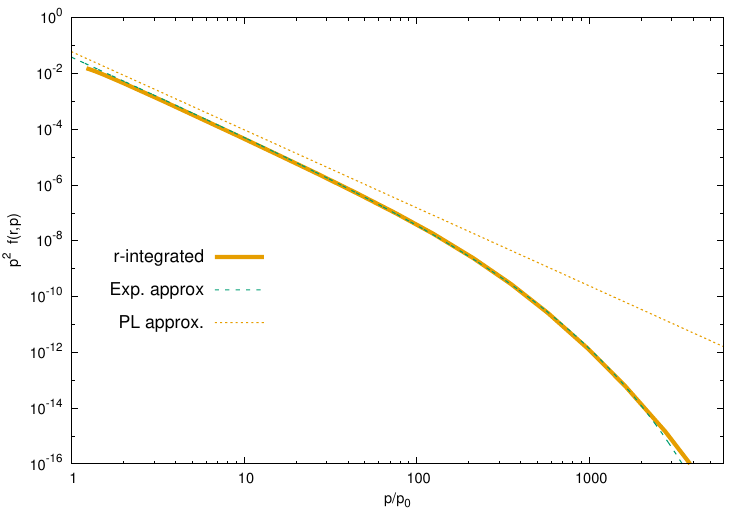}
\caption{The spatially-integrated electron momentum distribution $n(p) \propto p^2 \int dr 
\, r\, f(r,p)$ for the linearly decreasing flow profile, equation~(\ref{linear}), with 
$\beta_0 = 0.95$. A spatially constant $\tau \propto p^{\alpha}$ with $\alpha  =1/3$, and 
$\gamma_0=10^6$ have been assumed. The thin dotted line gives the power-law ($f(p) 
\propto p^{-s}$) solution ($s\simeq 4.8$) in the absence of synchrotron losses 
\citep{Rieger2022}, the dashed line shows a numerical fit to the exponential cut-off shape, 
$\exp(-0.05\,[p/p_0]^{2/3})$.}
\label{fig3}
\end{center}
\end{figure}

Figure~\ref{fig1} shows examples of the electron distribution $n(r,p) 
\propto p^2 f(r,p)$ above injection, evaluated at different spatial locations 
$x =r/r_2$ for a linearly deceasing flow profile with trans-relativistic on-axis 
speed $\beta_0 = 0.7$.
In the cut-off region the particle distribution is generally well described by 
a sub-exponential dependence of the form equation~(\ref{shape}). Since in this 
case the flow is trans-relativistic with $\gamma_b(r) \leq 1.4$, the synchrotron 
cut-offs resemble each other. For higher flow speeds, the situation changes. 
This is illustrated in Figure~\ref{fig2} for $\beta_0=0.95$ ($\gamma_{b,0}
= 3.2$). Since the flow Lorentz factor decreases with $x$, higher 
momenta are thus more readily achieved at smaller $x$. To facilitate comparison 
with a box treatment, we also calculate the spatially averaged distribution $n(p) 
\propto p^2 \int_0^{r_2} r f(r,p) dr$ in Figure~\ref{fig3}.
Comparing the approximation (thin red-dotted line) for the exponential shape in 
Figure~\ref{fig1} with equation~(\ref{shape}), suggests a reference value 
$\gamma_{\rm max} \simeq 600 \,\gamma_0 \simeq 6\times 10^7$ for this low, 
$\beta_0=0.7$ case. On the other hand, for the high $\beta_0=0.95$ case a 
comparison with the approximation (dashed) of the exponential shape 
in Figure~\ref{fig3}, indicates a reference value $\gamma_{\rm max} \simeq 
1.5\times10^3 \,\gamma_0 \sim 1.5 \times 10^9$. 
%

\subsection{Numerical solutions for power-law profiles}
\begin{figure}[htbp]
\begin{center}
\includegraphics[width = 0.48 \textwidth]{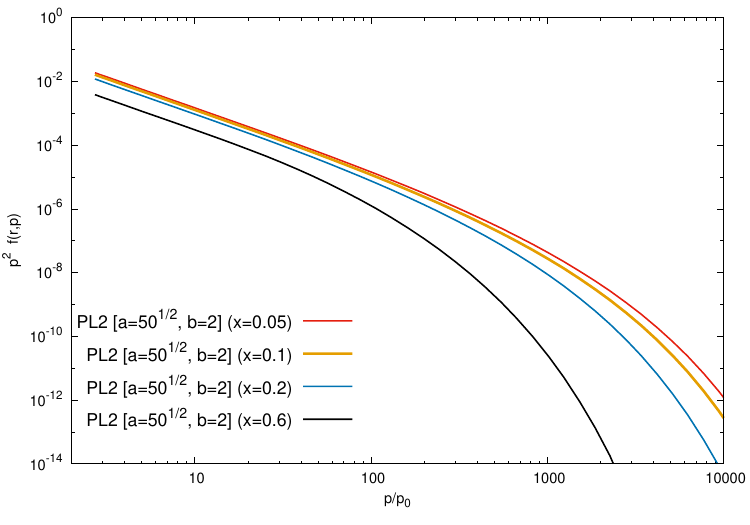}
\caption{The accelerated electron momentum distribution $n(r,p) \propto p^2 f(r,p)$ 
for a power-law flow profile, equation~(\ref{power-law}), with $a_p=\sqrt{50}$, $b= 2$ 
and $\beta_0=0.95$, evaluated at different radii $r=x\,r_2$. A spatially constant 
$\tau \propto p^{\alpha}$ with $\alpha=1/3$ and $\gamma_0 = 10^6$, along with 
$B=20\mu$G, have been assumed. The distribution extends to higher momenta at 
smaller radii where faster flow speeds ($\gamma_b > 1$) are met.}
\label{fig4}
\end{center}
\end{figure}

At mildly relativistic speeds, power-law decreasing flow profiles can lead to harder 
particle spectra \citep{Rieger2022}. Figure~\ref{fig4} shows examples of the resultant 
electron distribution $n(r,p) \propto p^2 f(r,p)$ above injection, evaluated at different 
spatial locations for a power-law flow profile with relativistic on-axis speed 
$\beta_0 = 0.95$. The (lower-energy) power-law part of the spectrum is 
harder ($s \simeq 3.9$) as compared to the respective ($\beta_0=0.95$), linearly 
decreasing profile (where $s\simeq 4.8$), cf. Figure~\ref{fig3}. Correspondingly, 
higher maximum Lorentz factors $\gamma_{\rm max}$ are achieved (note that a 
magnetic field twice as strong has been employed).

\subsection{Acceleration timescales and cut-off energies}
\begin{figure}[tbp]
\begin{center}
\includegraphics[width = 0.48 \textwidth]{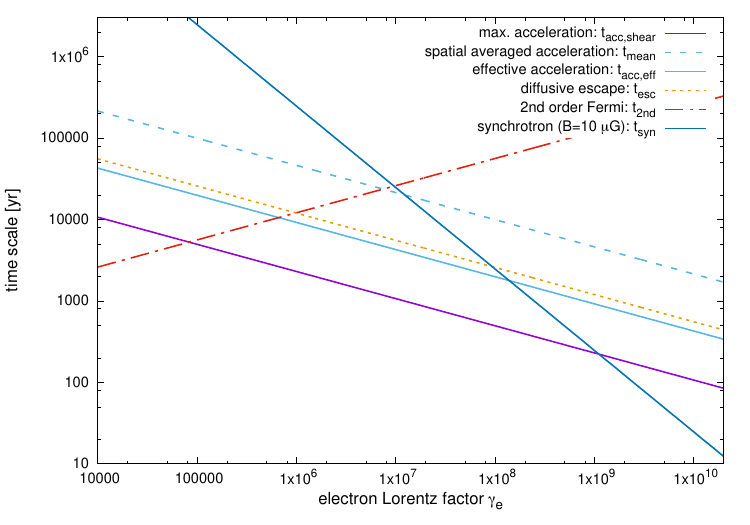}
\caption{Illustrative example of the timescales under consideration. 
A mildly relativistic flow with a linearly decreasing profile with 
spine Lorentz factor $\gamma_{b,0} =2$, Kolmogorov turbulence $\alpha
=1/3$, coherence length $\Lambda_{\rm max} = \Delta r = 0.1$ kpc, $\xi=0.2$ 
and Alfven speed $\beta_A=1/30$, along with $B=10\mu$G has been assumed. 
The pink solid curve gives the fastest acceleration $t_{\rm acc,shear}$, 
cf. eq.~(\ref{fastest}), while the blue solid line denotes the effective
acceleration time $t_{\rm acc,eff}$, cf. equation~(\ref{effective}). 
In this case shear acceleration starts to dominates over second-order 
Fermi acceleration ($t_{\rm 2nd}$) above $\gamma_0 \sim 10^5$ and is 
limited to $\gamma_{\rm e,max} \sim 10^9$, cf. Equation~(\ref{local}), due 
to synchrotron losses ($t_{\rm syn}$).}
\label{fig5}
\end{center}
\end{figure}

In general, the acceleration efficiency in a shearing flow with velocity $\beta(r)$ 
depends on the spatial coordinate $r$ \citep[][]{Rieger2019}
\beq\label{fastest}
t_{\rm acc}(r) \propto \frac{1}{\gamma_b(r)^{4} (\partial \beta/\partial r)^{2}}\,.
\eeq 
Typically, for radially decreasing, relativistic flow profiles the fastest 
acceleration of order $t_{\rm acc} \propto 1/\gamma_{b,0}^{4}$, with $\gamma_{b,0} 
\equiv (1-\beta_0^2)^{-1/2}>1$, is achieved on small spatial scales (see 
Figures~\ref{fig1},\ref{fig2} and \ref{fig4}), where the 
flow Lorentz factors (and flow gradients) are maximal, while acceleration becomes 
less efficient on larger scales when $\gamma_b(r) \rightarrow 1$. Balancing the 
local (comoving) acceleration time scale with the synchrotron cooling ($t_{\rm cool}
(\gamma)=9m^3c^5/[4e^4\gamma B^2]$) then implies that for a uniform magnetic field 
$B$, higher electron Lorentz factors can be achieved near to the inner boundary of 
the jet shear, $\gamma_{\rm e,max} \propto \gamma_{b,0}^6$. 
For instance, in the case of a linearly decreasing profile
\beq\label{local}
\gamma_{\rm e, max} \simeq 1.4\times 10^9 \,\xi_2^{-3/2}
                    \left(\frac{\gamma_{b,0}}{2}\right)^6
                    \left(\frac{100\,{\rm pc}}{\Delta r}\right)^2
                   \left(\frac{10\, \mu{\rm G}}{B}\right)^{7/2}\,,
\eeq with $\xi_2 =(\xi/0.2)$ and $B$ the magnetic field strength measured in the 
jet frame \citep[][]{Rieger2019b}, as is shown in  Fig.~\ref{fig5}.

In principle, a comparison of the particle spectrum with equation~(\ref{s_LB}) of 
the box model can also be used to infer a mean (spatially averaged) acceleration 
timescale \citep[][]{Rieger2019,Wang2021}
\beq\label{mean}
\bar{t}_{\rm acc} = w \tau_{\rm esc}/(4+\alpha)\,,
\eeq where $w$ encapsulates the dependency on the flow profile. In particular, for 
a linearly decreasing flow profile \citep{Rieger2022}
\beq
 w \simeq 116 \left( \ln\left[\frac{1+\beta_{02}}{1-\beta_{02}}\right]\right)^{-2}\,,
\eeq where $\beta_{02} = (\beta_0-\beta_2)/(1-\beta_0\beta_2)$ denotes the
relativistic relative velocity and $\beta_2 = \beta(r_2)$. 
Balancing $\bar{t}_{\rm acc}$ with the synchrotron cooling timescale gives a 
conservative lower bound on achievable particle energies,
\beq
\bar{\gamma}_{\rm e} \simeq 8.4 \times 10^8 w^{-3/2}\,\xi_2^{-3/2}
                    \left(\frac{100\,{\rm pc}}{\Delta r}\right)^2
                    \left(\frac{10\, \mu{\rm G}}{B}\right)^{7/2}\,.
\eeq For the linearly decreasing examples explored above (Sec.~\ref{subsec31}), 
one finds $\bar{\gamma}_{\rm e}\simeq 3.5\times 10^6$ ($\beta_0 =0.7$) and 
$\bar{\gamma}_{\rm e} \simeq 3.3 \times 10^7$ ($\beta_0=0.95$), respectively.
In reality, however, we expect that in relativistic flows higher electron energies 
are achieved (equation~(\ref{local})) as higher flow speeds than the mean are 
experienced. The extent to which this becomes noticeable, will depend on the 
topology and orientation of the flow. We can roughly explore this in our case by 
integrating $f(r,p) \propto \exp[-6.5\, (\gamma/\gamma_{\rm max}(r))^{2/3}]$ 
(equation~(\ref{shape})) over the cross-sectional area of the jet; i.e. evaluating 
$2\pi \int r f(r,p) dr$ with $\gamma_{\rm max}(r) = \gamma_* \gamma_b(r)^6$, 
where $\gamma_* = \gamma_{\rm e,max}/\gamma_{\rm b,0}^6$. Taking the 
asymptotic expansion of the resultant imaginary error function reproduces the 
exponential shape of equation~(\ref{shape}), but with $\gamma_{\rm max}(r)$ 
now replaced by an effective cut-off Lorentz factor $\gamma_{\rm max}^{\rm eff}$ 
given by $\gamma_{\rm max}^{\rm eff} \simeq \gamma_{\rm b,0}^3  \gamma_* 
\simeq \gamma_{\rm e,max}/\gamma_{\rm b,0}^3$. Accordingly,
\beq\label{effective}
  \gamma_{\rm max}^{\rm eff}\simeq 2.4\times 10^8 \,\xi_2^{-3/2}
                        \left(\frac{\gamma_{b,0}}{2}\right)^3
                        \left(\frac{100\,{\rm pc}}{\Delta r}\right)^2
                        \left(\frac{10\, \mu{\rm G}}{B}\right)^{7/2}\,.
\eeq For the linearly decreasing examples studied above (Sec.~\ref{subsec31}), 
this yields $\gamma_{\rm max}^{\rm eff}\simeq 8\times 10^7$ ($\beta_0 =0.7$) and 
$\gamma_{\rm max}^{\rm eff}\simeq 10^9$ ($\beta_0=0.95$)  respectively. This is 
in very good agreement with the results found in Sec.~\ref{subsec31} and suggests 
that for mildly relativistic flows $\gamma_{\rm max}^{\rm eff}$ provides a 
suitable, first-order approximation to the electron cut-off energies achievable 
in a phenomenological box model approach. 

In general, at higher flow speeds and for spatially varying magnetic fields, the 
situation is more complex and explicit modelling will be required. 
For instance, in the case of synchrotron emission, an exponential cut-off in the 
particle momentum distribution $n(\gamma) \propto \exp[-(\gamma/\gamma_c)^{\eta}]$, 
$\eta>0$, will be transformed into a synchrotron spectrum which is much smoother, 
$j_{\rm \nu} \propto \exp[-(\nu/\nu_c)^{\eta/(2+\eta}]$ 
\citep{Fritz1989,Zirakashvili2007}. In our case, with $\eta =2/3$, this would be changed 
into a rather weak sub-exponential cutoff with index $\eta/(2+\eta)=1/4$. This is actually 
comparable to the spectral shape expected for (non-relativistic, synchrotron-limited) shock 
acceleration in the presence of energy-independent diffusion \citep[][]{Zirakashvili2007}. 
A distant observer will see the jet emission differentially boosted, with observed 
flux approximately proportional to $2\pi \int r\,D(r)^2 j_{\nu}(r) dr$ where $D(r) 
= 1/[\gamma_b(r)(1-\beta(r)\cos i)]$ is the Doppler factor. For an aligned ($i=0$)
observer $D(r) =[1+\beta(r)]\gamma_b(r)$. Expanding the argument of the exponential
function in $j_{\nu}$ to first order in $r$ and performing the integration indicates
that for a relativistic jet viewed face-on, the cut-off frequency in the observed 
spectrum can reach up to $\nu_c\propto \gamma_{\rm e,max}^2 B$. Obviously then, 
given the sub-exponential shape and strong Doppler boosting effects, the observed 
synchrotron emission may extend to energies beyond those related to 
$\gamma_{\rm e,max}$ as expressed in equation~(\ref{local}). It is clear that a more 
complex situation arises for a non-uniform scattering time; we leave its detailed 
analysis to future work. In general, particle acceleration in shearing AGN jets 
is also a prime candidate mechanism for the production of ultra-high-energy 
cosmic-rays (UHECRs)\citep[e.g.,][]{Rieger2019b,Wang2021,Webb2023}. 
In the case of Centaurus A, for example, recent particle-MHD simulations illustrate 
that Hillas-type particle energies can plausibly be reached within its mildly 
relativistic large-scale jet \citep{Wang2024}. For completeness, considering UHECR 
acceleration, one may want to incorporate dynamical escape (along the jet) into the 
escape term of the transport equations~(\ref{diffusion}) and (\ref{pte}). Since the 
effective acceleration timescale derived here is shorter than the spatially averaged 
timescale, eq.~(\ref{mean}), conclusions regarding the UHECR potential of large-scale 
AGN jets as discussed in, e.g., \citet{Wang2021} are reinforced.

\section{Conclusions}
In many situations, radiative losses are expected to constrain the efficiency of
electron acceleration in jetted AGNs. Exploring a basic shear particle acceleration 
model and considering a conventional set of parameters, we have shown that in the 
presence of synchrotron losses electrons may well reach energies beyond 50 TeV in 
mildly relativistic large-scale AGN jets, providing a rationale for the detection of 
extended high-energy emission and the apparent presence of a second population of 
X-ray synchrotron-emitting particles \cite[][]{Sun2018,HESS2020,Perlman2020,
Breiding2023,He2023}. In general, efficient shear acceleration relies on seed 
injection of energetic electrons. Under typical conditions, those could well be 
provided by classical second-order Fermi or diffusive shock-type particle 
acceleration \citep[][]{Liu2017}; see Fig.~\ref{fig5}. Along with the fact 
that the highest electron energies are likely to be reached close to the inner shear 
boundary, this suggests that a shear model is phenomenologically better viewed as 
a multizone scenario; not necessarily requiring the observed radio and X-ray 
emission to be co-spatial. Such scenarios appear increasingly favored by 
a variety of observations \citep[][]{Kataoka2008,Reddy2021,He2023}. Inspection 
of Figures~\ref{fig2} and \ref{fig3} illustrates that a significant part of the X-ray 
emission could be associated with small ($r\sim$ few parsecs) spatial scales. We 
speculate that, together with differential Doppler boosting, this could potentially 
make it possible to accommodate year-type, few percent-level X-ray flux fluctuations for 
which evidence has been been recently reported \citep{Meyer2023}.

\acknowledgments
FMR kindly acknowledges funding by a DFG research grant under RI 1187/8-1. 
Discussions with Felix Aharonian, Martin Lemoine and Jieshuang Wang are 
gratefully acknowledged. 
 
\bibliography{references}{}

\begin{thebibliography}{}
\expandafter\ifx\csname natexlab\endcsname\relax\def\natexlab#1{#1}\fi
\providecommand{\url}[1]{\href{#1}{#1}}
\providecommand{\dodoi}[1]{doi:~\href{http://doi.org/#1}{\nolinkurl{#1}}}
\providecommand{\doeprint}[1]{\href{http://ascl.net/#1}{\nolinkurl{http://ascl.net/#1}}}
\providecommand{\doarXiv}[1]{\href{https://arxiv.org/abs/#1}{\nolinkurl{https://arxiv.org/abs/#1}}}

\bibitem[{{Berezhko}(1982)}]{Berezhko1982}
{Berezhko}, E.~G. 1982, Soviet Astronomy Letters, 8, 403

\bibitem[{{Berezhko} \& {Krymskii}(1981)}]{Berezhko1981}
{Berezhko}, E.~G., \& {Krymskii}, G.~F. 1981, Soviet Astronomy Letters, 7, 352

\bibitem[{{Breiding} {et~al.}(2023){Breiding}, {Meyer}, {Georganopoulos},
  {Reddy}, {Kollmann}, \& {Roychowdhury}}]{Breiding2023}
{Breiding}, P., {Meyer}, E.~T., {Georganopoulos}, M., {et~al.} 2023, \mnras,
  518, 3222, \dodoi{10.1093/mnras/stac3081}

\bibitem[{{Fritz}(1989)}]{Fritz1989}
{Fritz}, K.~D. 1989, \aap, 214, 14

\bibitem[{{He} {et~al.}(2023){He}, {Sun}, {Wang}, {Rieger}, {Liu}, \&
  {Liang}}]{He2023}
{He}, J.-C., {Sun}, X.-N., {Wang}, J.-S., {et~al.} 2023, \mnras, 525, 5298,
  \dodoi{10.1093/mnras/stad2542}

\bibitem[{{H.E.S.S. Collaboration} {et~al.}(2020){H.E.S.S. Collaboration},
  {Abdalla}, {Adam}, {Aharonian}, {Ait Benkhali}, {Ang{\"u}ner}, {Arakawa},
  {Arcaro}, {Armand}, {Ashkar}, {Backes}, {Barbosa Martins}, {Barnard},
  {Becherini}, {Berge}, {Bernl{\"o}hr}, {Blackwell}, {B{\"o}ttcher}, {Boisson},
  {Bolmont}, {Bonnefoy}, {Bregeon}, {Breuhaus}, {Brun}, {Brun}, {Bryan},
  {B{\"u}chele}, {Bulik}, {Bylund}, {Capasso}, {Caroff}, {Carosi}, {Casanova},
  {Cerruti}, {Chand}, {Chandra}, {Chen}, {Colafrancesco}, {Cury{\l}o},
  {Davids}, {Deil}, {Devin}, {deWilt}, {Dirson}, {Djannati-Ata{\"\i}},
  {Dmytriiev}, {Donath}, {Doroshenko}, {Drury}, {Dyks}, {Egberts}, {Emery},
  {Ernenwein}, {Eschbach}, {Feijen}, {Fegan}, {Fiasson}, {Fontaine}, {Funk},
  {F{\"u}{\ss}ling}, {Gabici}, {Gallant}, {Gat{\'e}}, {Giavitto}, {Glawion},
  {Glicenstein}, {Gottschall}, {Grondin}, {Hahn}, {Haupt}, {Heinzelmann},
  {Henri}, {Hermann}, {Hinton}, {Hofmann}, {Hoischen}, {Holch}, {Holler},
  {Horns}, {Huber}, {Iwasaki}, {Jamrozy}, {Jankowsky}, {Jankowsky},
  {Jardin-Blicq}, {Jung-Richardt}, {Kastendieck}, {Katarzy{\'n}ski},
  {Katsuragawa}, {Katz}, {Khangulyan}, {Kh{\'e}lifi}, {King}, {Klepser},
  {Klu{\'z}niak}, {Komin}, {Kosack}, {Kostunin}, {Kraus}, {Lamanna}, {Lau},
  {Lemi{\`e}re}, {Lemoine-Goumard}, {Lenain}, {Leser}, {Levy}, {Lohse},
  {Lypova}, {Mackey}, {Majumdar}, {Malyshev}, {Marandon}, {Marcowith}, {Mares},
  {Mariaud}, {Mart{\'\i}-Devesa}, {Marx}, {Maurin}, {Meintjes}, {Mitchell},
  {Moderski}, {Mohamed}, {Mohrmann}, {Moore}, {Moulin}, {Muller}, {Murach},
  {Nakashima}, {de Naurois}, {Ndiyavala}, {Niederwanger}, {Niemiec}, {Oakes},
  {O'Brien}, {Odaka}, {Ohm}, {de Ona Wilhelmi}, {Ostrowski}, {Oya}, {Panter},
  {Parsons}, {Perennes}, {Petrucci}, {Peyaud}, {Piel}, {Pita}, {Poireau},
  {Priyana Noel}, {Prokhorov}, {Prokoph}, {P{\"u}hlhofer}, {Punch},
  {Quirrenbach}, {Raab}, {Rauth}, {Reimer}, {Reimer}, {Remy}, {Renaud},
  {Rieger}, {Rinchiuso}, {Romoli}, {Rowell}, {Rudak}, {Ruiz-Velasco},
  {Sahakian}, {Saito}, {Sanchez}, {Santangelo}, {Sasaki}, {Schlickeiser},
  {Sch{\"u}ssler}, {Schulz}, {Schutte}, {Schwanke}, {Schwemmer},
  {Seglar-Arroyo}, {Senniappan}, {Seyffert}, {Shafi}, {Shiningayamwe},
  {Simoni}, {Sinha}, {Sol}, {Specovius}, {Spir-Jacob}, {Stawarz}, {Steenkamp},
  {Stegmann}, {Steppa}, {Takahashi}, {Tavernier}, {Taylor}, {Terrier},
  {Tiziani}, {Tluczykont}, {Trichard}, {Tsirou}, {Tsuji}, {Tuffs}, {Uchiyama},
  {van der Walt}, {van Eldik}, {van Rensburg}, {van Soelen}, {Vasileiadis},
  {Veh}, {Venter}, {Vincent}, {Vink}, {Voisin}, {V{\"o}lk}, {Vuillaume},
  {Wadiasingh}, {Wagner}, {White}, {Wierzcholska}, {Yang}, {Yoneda},
  {Zacharias}, {Zanin}, {Zdziarski}, {Zech}, {Ziegler}, {Zorn}, \&
  {{\.Z}ywucka}}]{HESS2020}
{H.E.S.S. Collaboration}, {Abdalla}, H., {Adam}, R., {et~al.} 2020, \nat, 582,
  356, \dodoi{10.1038/s41586-020-2354-1}

\bibitem[{{Jokipii} \& {Morfill}(1990)}]{Jokipii1990}
{Jokipii}, J.~R., \& {Morfill}, G.~E. 1990, \apj, 356, 255,
  \dodoi{10.1086/168837}

\bibitem[{{Kataoka} {et~al.}(2008){Kataoka}, {Stawarz}, {Harris},
  {Siemiginowska}, {Ostrowski}, {Swain}, {Hardcastle}, {Goodger}, {Iwasawa}, \&
  {Edwards}}]{Kataoka2008}
{Kataoka}, J., {Stawarz}, {\L}., {Harris}, D.~E., {et~al.} 2008, \apj, 685,
  839, \dodoi{10.1086/591024}

\bibitem[{{Kirk} {et~al.}(1988){Kirk}, {Schlickeiser}, \&
  {Schneider}}]{Kirk1988}
{Kirk}, J.~G., {Schlickeiser}, R., \& {Schneider}, P. 1988, \apj, 328, 269,
  \dodoi{10.1086/166290}

\bibitem[{{Lemoine}(2019)}]{Lemoine2019}
{Lemoine}, M. 2019, \prd, 99, 083006, \dodoi{10.1103/PhysRevD.99.083006}

\bibitem[{{Liu} {et~al.}(2017){Liu}, {Rieger}, \& {Aharonian}}]{Liu2017}
{Liu}, R.-Y., {Rieger}, F.~M., \& {Aharonian}, F.~A. 2017, ApJ, 842, 39,
  \dodoi{10.3847/1538-4357/aa7410}

\bibitem[{{Merten} {et~al.}(2021){Merten}, {Boughelilba}, {Reimer}, {Da Vela},
  {Vorobiov}, {Tavecchio}, {Bonnoli}, {Lundquist}, \& {Righi}}]{Merten2021}
{Merten}, L., {Boughelilba}, M., {Reimer}, A., {et~al.} 2021, Astroparticle
  Physics, 128, 102564, \dodoi{10.1016/j.astropartphys.2021.102564}

\bibitem[{{Meyer} {et~al.}(2023{\natexlab{a}}){Meyer}, {Shaik}, {Reddy}, \&
  {Georganopoulos}}]{Meyer2023IAUS}
{Meyer}, E., {Shaik}, A., {Reddy}, K., \& {Georganopoulos}, M.
  2023{\natexlab{a}}, in IAU Symposium, Vol. 375, The Multimessenger Chakra of
  Blazar Jets, ed. I.~{Liodakis}, M.~F. {Aller}, H.~{Krawczynski},
  A.~{L{\"a}hteenm{\"a}ki}, \& T.~J. {Pearson}, 1--8,
  \dodoi{10.1017/S1743921323000911}

\bibitem[{{Meyer} {et~al.}(2023{\natexlab{b}}){Meyer}, {Shaik}, {Tang}, {Reid},
  {Reddy}, {Breiding}, {Georganopoulos}, {Chiaberge}, {Perlman}, {Clautice},
  {Sparks}, {DeNigris}, \& {Trevor}}]{Meyer2023}
{Meyer}, E.~T., {Shaik}, A., {Tang}, Y., {et~al.} 2023{\natexlab{b}}, Nature
  Astronomy, \dodoi{10.1038/s41550-023-01983-1}

\bibitem[{{Perlman} {et~al.}(2020){Perlman}, {Clautice}, {Avachat}, {Cara},
  {Sparks}, {Georganopoulos}, \& {Meyer}}]{Perlman2020}
{Perlman}, E.~S., {Clautice}, D., {Avachat}, S., {et~al.} 2020, Galaxies, 8,
  71, \dodoi{10.3390/galaxies8040071}

\bibitem[{{Reddy} {et~al.}(2021){Reddy}, {Georganopoulos}, \&
  {Meyer}}]{Reddy2021}
{Reddy}, K., {Georganopoulos}, M., \& {Meyer}, E.~T. 2021, \apjs, 253, 37,
  \dodoi{10.3847/1538-4365/abd8d7}

\bibitem[{{Rieger}(2019)}]{Rieger2019}
{Rieger}, F.~M. 2019, Galaxies, 7, 3.
\newblock \doarXiv{1909.07237}

\bibitem[{{Rieger} \& {Duffy}(2006)}]{Rieger2006}
{Rieger}, F.~M., \& {Duffy}, P. 2006, \apj, 652, 1044, \dodoi{10.1086/508056}

\bibitem[{{Rieger} \& {Duffy}(2019)}]{Rieger2019b}
---. 2019, ApJL, 886, L26, \dodoi{10.3847/2041-8213/ab563f}

\bibitem[{{Rieger} \& {Duffy}(2021)}]{Rieger2021}
---. 2021, \apjl, 907, L2, \dodoi{10.3847/2041-8213/abd567}

\bibitem[{{Rieger} \& {Duffy}(2022)}]{Rieger2022}
---. 2022, \apj, 933, 149, \dodoi{10.3847/1538-4357/ac729c}

\bibitem[{{Seo} {et~al.}(2023){Seo}, {Ryu}, \& {Kang}}]{Seo2023}
{Seo}, J., {Ryu}, D., \& {Kang}, H. 2023, \apj, 944, 199,
  \dodoi{10.3847/1538-4357/acb3ba}

\bibitem[{{Sun} {et~al.}(2018){Sun}, {Yang}, {Rieger}, {Liu}, \&
  {Aharonian}}]{Sun2018}
{Sun}, X.-N., {Yang}, R.-Z., {Rieger}, F.~M., {Liu}, R.-Y., \& {Aharonian}, F.
  2018, \aap, 612, A106, \dodoi{10.1051/0004-6361/201731716}

\bibitem[{{Tavecchio}(2021)}]{Tavecchio2021}
{Tavecchio}, F. 2021, \mnras, 501, 6199, \dodoi{10.1093/mnras/staa4009}

\bibitem[{{Wang} {et~al.}(2021){Wang}, {Reville}, {Liu}, {Rieger}, \&
  {Aharonian}}]{Wang2021}
{Wang}, J.-S., {Reville}, B., {Liu}, R.-Y., {Rieger}, F.~M., \& {Aharonian},
  F.~A. 2021, \mnras, 505, 1334, \dodoi{10.1093/mnras/stab1458}

\bibitem[{{Wang} {et~al.}(2023){Wang}, {Reville}, {Mizuno}, {Rieger}, \&
  {Aharonian}}]{Wang2023}
{Wang}, J.-S., {Reville}, B., {Mizuno}, Y., {Rieger}, F.~M., \& {Aharonian},
  F.~A. 2023, \mnras, 519, 1872, \dodoi{10.1093/mnras/stac3616}

\bibitem[{{Wang} {et~al.}(2024){Wang}, {Reville}, {Rieger}, \&
  {Aharonian}}]{Wang2024}
{Wang}, J.-S., {Reville}, B., {Rieger}, F.~M., \& {Aharonian}, F.~A. 2024,
  \apjl, 977, L20, \dodoi{10.3847/2041-8213/ad9589}

\bibitem[{{Webb}(1989)}]{Webb1989}
{Webb}, G.~M. 1989, ApJ, 340, 1112, \dodoi{10.1086/167462}

\bibitem[{{Webb} {et~al.}(2019){Webb}, {Al-Nussirat}, {Mostafavi}, {Barghouty},
  {Li}, {le Roux}, \& {Zank}}]{Webb2019}
{Webb}, G.~M., {Al-Nussirat}, S., {Mostafavi}, P., {et~al.} 2019, \apj, 881,
  123, \dodoi{10.3847/1538-4357/ab2fca}

\bibitem[{{Webb} {et~al.}(2018){Webb}, {Barghouty}, {Hu}, \& {le
  Roux}}]{Webb2018}
{Webb}, G.~M., {Barghouty}, A.~F., {Hu}, Q., \& {le Roux}, J.~A. 2018, \apj,
  855, 31, \dodoi{10.3847/1538-4357/aaae6c}

\bibitem[{{Webb} {et~al.}(2023){Webb}, {Xu}, {Biermann}, {Al-Nussirat},
  {Mostafavi}, {Li}, {Barghouty}, \& {Zank}}]{Webb2023}
{Webb}, G.~M., {Xu}, Y., {Biermann}, P.~L., {et~al.} 2023, \apj, 958, 169,
  \dodoi{10.3847/1538-4357/acfda9}

\bibitem[{{Zirakashvili} \& {Aharonian}(2007)}]{Zirakashvili2007}
{Zirakashvili}, V.~N., \& {Aharonian}, F. 2007, \aap, 465, 695,
  \dodoi{10.1051/0004-6361:20066494}

\end{thebibliography}
\bibliographystyle{aasjournal}

\end{document}